\documentstyle[12pt]{article}
\input{epsfig.sty}
\newcommand{\beq}{\begin{eqnarray}}
\newcommand{\eeq}{\end{eqnarray}}
\begin{document}
\title{B Production In p-p and A-A Collisions}
\author{Leonard S. Kisslinger and Bijit Singha\\
Department of Physics, Carnegie Mellon University, Pittsburgh, PA 15213}
\date{}
\maketitle
\begin{abstract}
This is an extension of our recent work on $D^+(c\bar{d}),D^o(c\bar{u})$ 
production from  p-p and d-Au collisions to $B^+(b\bar{d}),B^o(b\bar{u})$
production from  p-p and A-A collisions. The rapidity cross sections for
$B^+(c\bar{d}),B^o(b\bar{u})$ production from both p-p and A-A collisions
are estimated. Our present work makes use of previous work on $J/\Psi$, 
$\Psi'(2S)$, $\Upsilon(nS)$ production in p-p and A-A collisions, with the main
new aspect being the fragmentation probability, $D_{b \rightarrow b\bar{q}}$, which 
turns out to be similar to the fragmentation probability $D_{c \rightarrow c\bar{q}}$
used in our recent work.
\end{abstract}
\noindent
PACS Indices:12.38.Aw,13.60.Le,14.40.Lb,14.40Nd
\vspace{1mm}

\section{Introduction}
  We consider $B^+(b\bar{d}),B^o(b\bar{u})$ production via unpolarized p-p 
collisions at 200 GeV, an extension of our recent work on $D^+(c\bar{d}),
D^o(c\bar{u})$ production\cite{klm17}. We make use of previous work on 
$J/\Psi,\Psi'(2S)$ and $\Upsilon(nS)$ productiona via p-p collisions\cite{klm11}
and A-A collisions\cite{klm14}.  In addition to being an important 
study of QCD, the estimate of $B$ production via A-A collisions could provide 
a test of the production of Quark Gluon Plasma (QGP) in relativistic heavy ion 
collisions (RHIC),

  As in our previous work we use the color octet model\cite{cl96,bc96,fl96},
which is consistent with experimental studies at E=200 GeV \cite{nlc03,cln04}. 
In Refs.\cite{klm11},\cite{klm14} the mixed hybrid theory was used for the
production of $\Psi'(2S),\Upsilon(3S)$, but this not relevant for the
present theory.

  The main new aspect of the present work is that while a gluon can produce a
$c\bar{c}$ or $b\bar{b}$ state, it cannot directly produce a $b\bar{d}$.
A fragmentation process converts a $b\bar{b}$ into a $b\bar{d}-d\bar{b}$,
for example. We use the fragmentation probability, $D_{b \rightarrow b\bar{q}}$ of 
Braaten et. al.\cite{bcy93,bcfy95}.

\section{Differential $pp\rightarrow BX$ cross section}

Using what in Ref\cite{ns06} is called scenerio 2, the production
cross section with gluon dominance for BX is
\beq
\label{1}
  \sigma_{pp\rightarrow BX} &=&  \int_a^1 \frac{d x}{x} 
f_g(x,2m)f_g(a/x,2m) \sigma_{gg\rightarrow BX}\; ,
\eeq
with\cite{bcfy95}
\beq
\label{2}
 \sigma_{gg\rightarrow BX}&=& 2 \sigma_{gg\rightarrow b\bar{b}}
D_{b\rightarrow b\bar{q}} \; ,
\eeq
where $\sigma_{gg\rightarrow b\bar{b}}$ is similar to the charmonium production
cross section in Ref\cite{klm11} and $D_{b\rightarrow b\bar{q}}$ is the total
fragmentation probability. 
For $E=\sqrt{s}$=200 GeV the gluon distribution funtion for bottomonium quarks
is
\beq
\label{fg}
 f_g(y) &=& 275.14 - 6167.6 x(y) + 36871.3 x(y)^2
\eeq

We use the quark fragmentation probability, 
$D_{b \rightarrow b\bar{q}}$ of Braaten et. al.\cite{bcy93, bcfy95}.

From Ref\cite{bcfy95}, using for the light quark mass=(up-mass+down-mass)/2=
3.5 MeV.
\beq
\label{3}
    D_{b \rightarrow b\bar{q}}&=& 9.21 \times 10^{5} \alpha_s^2 |R(0)|^2/\pi 
 \; ,
\eeq
in units of $(1/GeV^3)$, with $\alpha_s=.26$. For a 1S state
$|R(0)| ^2 = 4/(a_o)^3$. For a $b\bar{q}$ state, $(1/a_o)=m_q\simeq 3.5$ MeV. 
Therefore,
\beq
\label{R(0)}
   |R(0)| ^2&\simeq& 1.71 \times 10^{-7} {\rm \;\;(GeV)^3} \nonumber \\
    D_{b \rightarrow b\bar{q}}&\simeq& 3.39 \times 10^{-3} \; ,
\eeq
so $D_{b \rightarrow b\bar{q}} \simeq D_{c \rightarrow c\bar{q}}$\cite{klm17}.
 
The calculation of the cross section is similar to that in Ref\cite{klm11}.
\beq
\label{4}
  \frac{d\sigma_{pp\rightarrow BX}}{dy}&=& Abb*f_g(x(y),2m) f_g(a/x(y),2m) 
\frac{dx(y)}{dy} \frac{1}{x(y)} D_{b \rightarrow b\bar{q}} \; ,
\eeq
with rapidity $y$
\beq
\label{y(x)}
      y &=& \frac{1}{2} ln (\frac{E + p_z}{E-p_z}) \nonumber \\
        x(y) &=& 0.5 \left[\frac{m}{E}(\exp{y}-\exp{(-y)})+\sqrt{(\frac{m}{E}
(\exp{y}-\exp{(-y)}))^2 +4a}\right] \; , 
\eeq
where $Abb$ is the matrix element for bottomonium production\cite{klm11} 
modified by an effective mass $ms$: $ Abb= 7.9*10^{-4} (1.5/ms)^3$ nb.
For $ms$ = 5.0 GeV $Abb= 2.13*10^{-5}$ nb.

 From Eq(\ref{4}) we find $\frac{d\sigma_{pp\rightarrow BX}}{dy}$ shown in
the figure below, with $ms$=5.0 GeV.
\vspace{2cm}

\begin{figure}[ht]
\begin{center}
\epsfig{file=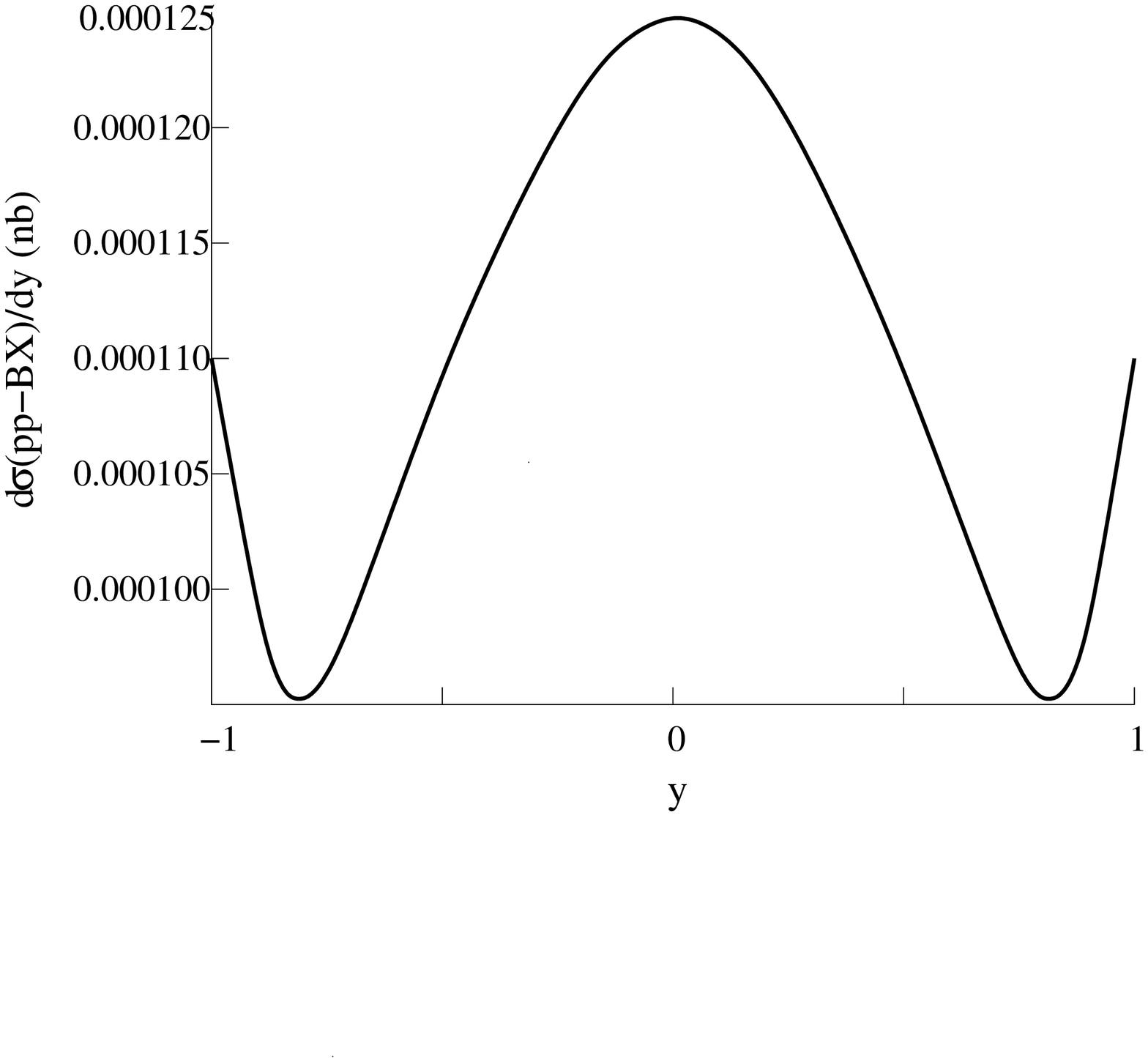,height=8cm,width=12cm}
\end{center}
\caption{$d\sigma/dy$ for E=200 GeV unpolarized p-p collisions producing B+X} 
\end{figure}

\section{Total $pp\rightarrow BX$ cross section}

The total cross section for $pp\rightarrow BX$ is\cite{klm11}
\beq
\label{5}
  \sigma_{pp\rightarrow BX}&=&  \int_a^1\frac{dx}{x} Abb*f_g(x(y),2m) 
f_g(a/x(y),2m)  D_{b \rightarrow b\bar{q}} \; .
\eeq

From Eqs(\ref{fg},\ref{R(0)}) and $Abb$ one obtains
\beq
\label{sigma}
  \sigma_{pp\rightarrow BX}&\simeq& 0.4823 {\rm nb} \; .
\eeq

Since $\sigma_{pp\rightarrow DX}\simeq 2680.0 {nb}$\cite{klm17}, the ratio of
$\sigma_{pp\rightarrow BX}$ to $\sigma_{pp\rightarrow DX}$ is
\beq
\label{RR}
 RR &\equiv& \frac{\sigma_{pp\rightarrow BX}}{\sigma_{pp\rightarrow DX}} \simeq
1.8\times10^{-4} \; ,
\eeq
due to the difference in the quark mass and values of $f_g$ for bottom vs
charm quarks.

\newpage

  A number of experiments have measured $\sigma_{b\bar{b}}$ cross 
sections at $\sqrt{s_{pp}}$=200 GeV\cite{phenix06,phenix07,star07,phenix09}.
Experimental measurements of $B^+, B^-, B^0$ production via p-p collisions 
are expected in the future.

\section{Differential $Cu-Cu, Au-Au \rightarrow BX$ cross sections}

Cold nuclear matter effects on heavy-quark production were estimated
for a number of rapidities via PHENIX experiments\cite{phenix14}. We use 
the results of this experiment for the study of $B$ production via Cu-Cu and 
Au-Au collisions.

In this Section we estimate the production of $B^+, B^0$ from Cu-Cu and Au-Au 
collisions, using the methods given in Ref.\cite{klm14} for the estimate of 
production of $\Psi$ and $\Upsilon$ states via Cu-Cu and Au-Au collisions
based on p-p collisions.

  The differential rapidity cross section for B+X production via A-A
collisions is given by $ \frac{d\sigma_{pp\rightarrow BX}}{dy}$ with modification
described in Ref.\cite{klm14} for Cu-Cu and Au-Au collisions:
\beq
\label{sigmadAu}
  \frac{d\sigma_{AA\rightarrow BX}}{dy}&=& R_{AA} N^{AA}_{bin}
\left (\frac{d\sigma_{pp\rightarrow BX}}{dy} \right)  \; ,
\eeq
$R_{AA}$  is the nuclear-modification factor, $N^{AA}_{bin}$ is the
number of binary collisions in the AA collision, and $\left
 (\frac{d\sigma_{pp\rightarrow BX}}{dy} \right)$ is the differential rapidity cross
section for $BX$ production via nucleon-nucleon collisions in the nuclear 
medium.

  $\left (\frac{d\sigma_{pp\rightarrow BX}}{dy} \right)$ is given by Eq(\ref{4})
with $x(y)$ replaced by the function $\bar{x}$, the effective parton x in the 
nucleus Au\cite{vitev06}:
\beq
\label{barx}
         \bar{x}(y)&=& x(y)(1+\frac{\xi_g^2(A^{1/3}-1)}{Q^2}) \; ,
\eeq
which was evaluated in Ref.\cite{klm14}, where it was shown that 
$\bar{x}(y) \simeq x(y)$

  Experimental studies show that for $\sqrt{s_{pp}}$ = 200 GeV 
$R^E_{AA}\simeq 0.5$ both for Cu-Cu\cite{star09,phenix08} and 
Au-Au\cite{phenix07,star07,kks06}. The
number of binary collisions are  $N^{AA}_{bin}$=51.5 for Cu-Cu\cite{sbstar07} 
and 258 for Au-Au.

  From Eqs(\ref{4}) and (\ref{sigmadAu}) one obtains the differential 
rapidity cross section for B+X production via Cu-Cu and Au-Au collisions
\beq
\label{sigCuCuAuAu}
    \frac{d\sigma_{Cu-Cu\rightarrow BX}}{dy}&=& (51.5/2)\times 
Abb*f_g(x(y),2m) f_g(a/x(y),2m) \frac{dx(y)}{dy} \frac{1}{x(y)} 
D_{b \rightarrow b\bar{q}} \nonumber \\
   \frac{d\sigma_{Au-Au\rightarrow BX}}{dy}&=& (258/2)\times 
Abb*f_g(x(y),2m) f_g(a/x(y),2m) \frac{dx(y)}{dy} \frac{1}{x(y)} 
D_{b \rightarrow b\bar{q}} \; .
\eeq

\newpage

  $d\sigma_{Cu-Cu\rightarrow BX}/dy$ and $d\sigma_{Au-Au\rightarrow BX}/dy$
are shown in the figures below.
 \vspace{5cm}
 
\begin{figure}[ht]
\begin{center}
\epsfig{file=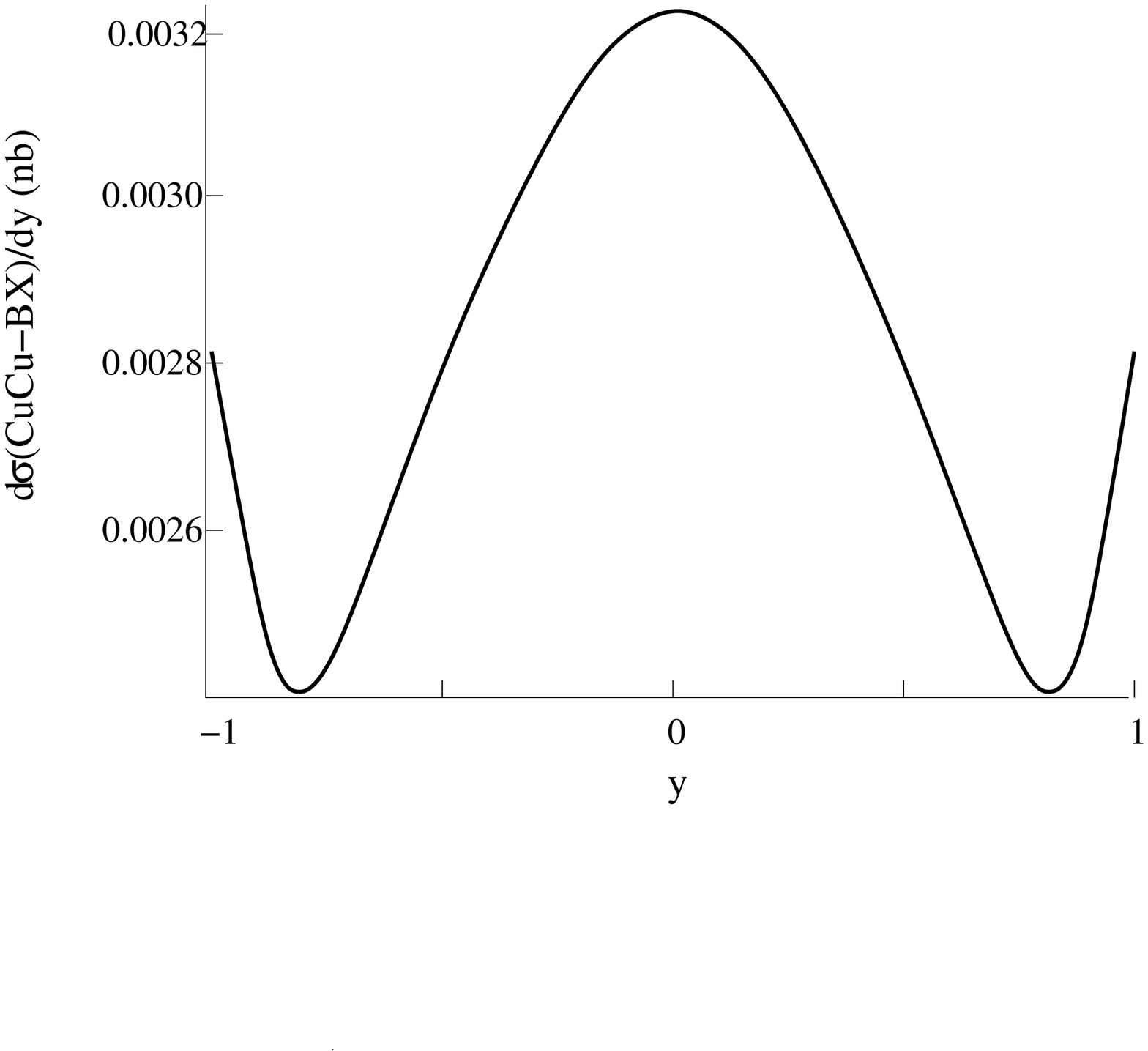,height=6cm,width=12cm}
\end{center}
\end{figure}
\vspace{-3.5cm}

\hspace{1cm}Figure 2 $d\sigma/dy$ for E=200 GeV Cu-Cu collisions producing B+X
\vspace{2cm}

\begin{figure}[ht]
\begin{center}
\epsfig{file=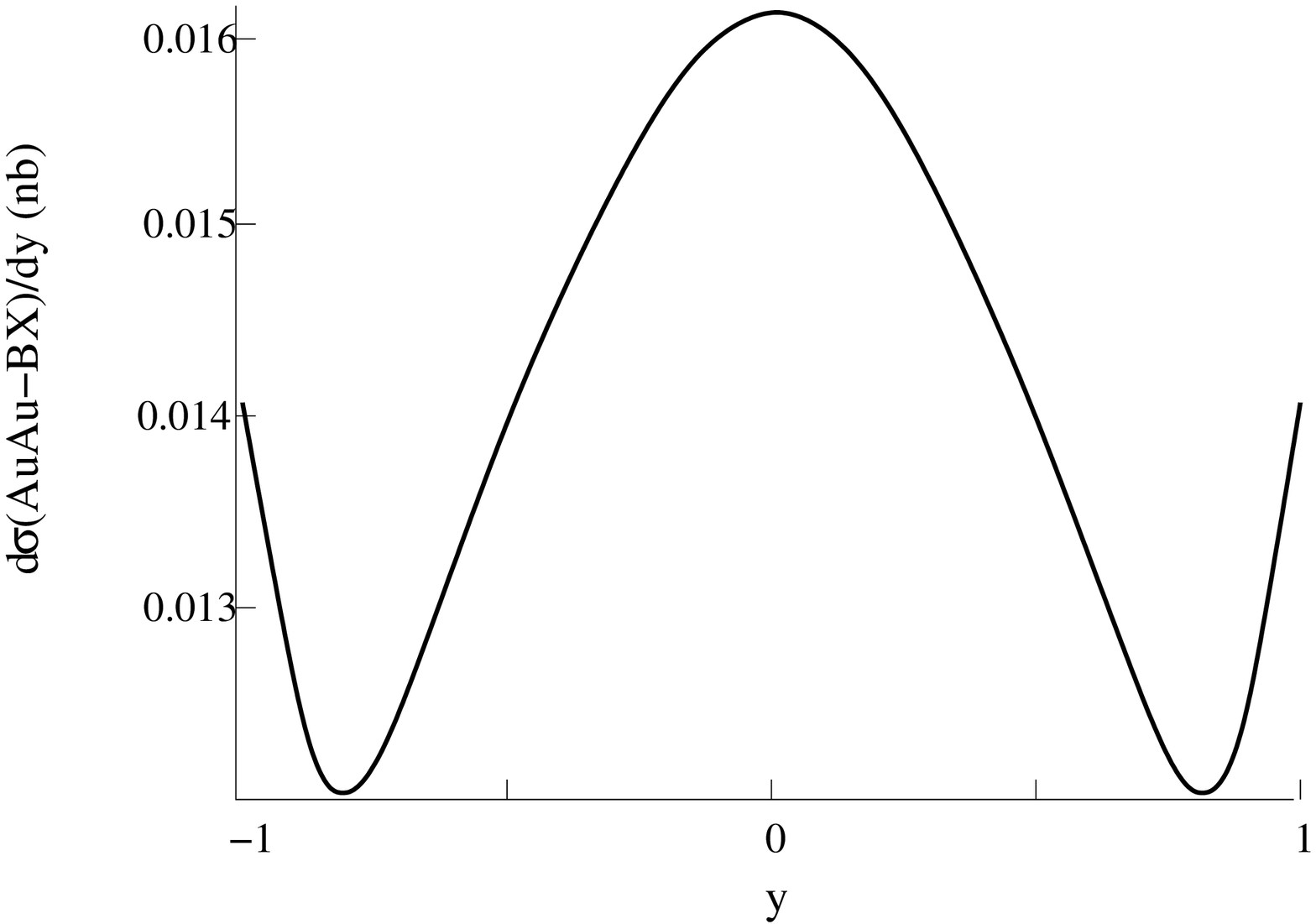,height=6cm,width=12cm}
\end{center}
\end{figure}

\hspace{1cm}Figure 3 $d\sigma/dy$ for E=200 GeV Au-Au collisions producing B+X
\newpage

\subsection{Total $CuCu\rightarrow BX$ and $AuAu\rightarrow BX$ 
cross sections}

  Total $\sigma_{CuCu\rightarrow BX}$ and $\sigma_{AuAu\rightarrow BX}$ crossections
are obtained from  $\sigma_{pp\rightarrow BX}$ Eq(\ref{sigma}) by multiplying
$\sigma_{pp\rightarrow BX}$ by $R_{AA} N^{AA}_{bin}$. Therefore
\beq
\label{CuCuAuAutotal}
  \sigma_{CuCu\rightarrow BX}&\simeq& (51.5/2)\times 0.4823 {\rm nb}=12.42
{\rm nb} \nonumber \\
   \sigma_{AuAu\rightarrow BX}&\simeq& (258/2)\times 0.4823 {\rm nb}=62.22 
{ \rm nb} \; .
\eeq

\section{Conclusions}

We have estimated the production of heavy-quark mesons 
$B^+(b\bar{d}),B^o(b\bar{u})$ +$X$ via p-p colllisions using the color
octet model with an extension of our previous work on production of
$\bar{c}c$ and $\bar{b}b$ states to $\bar{d}b$ or $\bar{u}b$ B-meson
states using fragmentation. Our results are expected to be tested by p-p 
collision experiments in the future. We have also estimated the production
of B-meson states via Cu-Cu and Au-Au collisions, using experimental results
for the nuclear modification and number of binary collisions in recent
A-A collisions experiments, which also might be measured in future
experiments. 

\vspace{1cm}
\Large{{\bf Acknowledgements}}\\
\normalsize
This work was supported in part by a grant from the Pittsburgh Foundation.

\end{document}